\documentstyle[prl,aps]{revtex}
\begin{document}

\title{\bf  Exact Quantum Monte Carlo Process for the Statistics
of Discrete Systems } 
\author{\ Nikolai V. Prokof'ev, Boris V. Svistunov, and Igor
S. Tupitsyn}
\address{
Russian Research Center ``Kurchatov Institute", 123182
Moscow, Russia.}

\maketitle
\begin{abstract}
We introduce an exact Monte Carlo approach to the statistics of
discrete quantum systems which does not rely on the standard fragmentation
of the imaginary time, or any small parameter. The method deals with
discrete objects, kinks, representing virtual transitions at different
moments of time. The global statistics of kinks is reproduced by explicit
local procedures, the key one being based on the exact solution for the
biased two-level system.
\end{abstract}


\bigskip

{\bf 1.} Quantum Monte Carlo (MC) simulation is the most powerful, if not
the only available method of getting  exact results for
complex systems \cite{A,B,C,D,E,F,G}, where analytic solutions are not
possible and exact diagonalization methods do not work because of the
enormous Hilbert space. One conventionally starts from the general
expression for the thermal average 
$ \langle \: A \: \rangle =   {\rm Tr}Ae^{-\beta H} /Z(\beta ) $,
where $Z(\beta )={\rm Tr}e^{-\beta H}$ is the partition function for the
Hamiltonian $H$, and converts the trace into the Feynman path integral 
\cite{Feynman}.
To sum over all 
trajectories with the weighting function $e^{-S}$ 
defined by the action $S$ the imaginary
time interval $[0,\beta ]$ is divided into $N_\beta =\beta /\Delta \tau $
smaller intervals of width $\Delta \tau $, and the actual trajectory is
described by specifying the system state $|\alpha _k\rangle $ at each time 
$\tau _k=k\Delta \tau $ where $k=0,1,...,N_\beta $.
To ensure summation over trajectories with the near-optimal
action, the complete sum is replaced by the stochastic sum (the
known Metropolis algorithm \cite{Metro}); in the most commonly used schemes
the probabilities $W_{1,2}$ to have the two trajectories $\{ \alpha_k
\}_{1,2}$ in the sum are related as 
\begin{equation}
W_1/W_2= e^{S_2-S_1} \; .  \label{3}
\end{equation}
After some trajectory is accepted to the statistics, the MC process
generates small variation to this trajectory which is either
accepted or rejected according to Eq.(\ref{3}).

An approximate treatment of noncommuting operators originating from the
artificial discretization of the time variable at points where the
trajectory changes, i.e., when $\vert \alpha_k \rangle \ne \vert
\alpha_{k+1} \rangle$ ( in what follows we call such points ``kinks"),
results in an intrinsic error $ \sim \Delta \tau ^2$
(see, e.g., \cite{Fye}). Taking smaller and smaller values of
$\Delta \tau$ eventually eliminates this intrinsic error, but then
the algorithm starts accumulating statistics too slowly.
Consider, as a typical example, the Hamiltonian of a particle on
a lattice 
\begin{equation}
H= -t\sum_{<ij>} a_i^{\dag}a_j + \sum_{i}U_{i}n_i \;,  \label{4}
\end{equation}
where $a_i^{\dag}$ creates a particle on site $i$, $t$ is the hopping matrix
element, $n_i=a_i^{\dag}a_i$, and $<ij>$ conventionally denotes nearest
neighbor sites. In this case the set of states $\{ \alpha \}$ may be chosen
to describe particle coordinate. The typical separation in time between two
adjacent kinks is of order $1/t$ and independent of $\Delta \tau$, so that
for small $\Delta \tau$ there are some $1/(t\Delta \tau ) \gg 1$ time
intervals between the kinks. Now, if the MC procedure
will suggest to create a new kink-antikink
pair then the corresponding variation of the trajectory will be most
probably rejected since the ratio $W_{new}/W_{old} \sim (t\Delta \tau )^2$
is proportional to the square of a small parameter. On another hand, when
the MC procedure suggests to shift the
already existing kink to the nearest point in time, the corresponding
variation of the trajectory is accepted with probability $\sim
O(1)$. Thus, in average, it takes some $1/(t\Delta \tau )^2$ attempts to
create a new kink - anti-kink pair, and $1/(t\Delta \tau)$ attempts to move
the kink to the nearest position in time. To conclude, the algorithm is
becoming progressively ineffective when $\Delta \tau $ is reduced.

In this paper we prove that for any Hamiltonian with discrete Hilbert space
it is possible to arrange an exact MC process, i.e., when both the
description of the trajectories in time and their actions contain no
approximations at all, and which is at least $1/(t\Delta \tau )^2$ times
faster than the conventional scheme in accumulating statistics.

First, we observe that for discrete systems like (\ref{4}) any trajectory
may be easily described using continuous time variable: it is sufficient to
define the state at $\tau =0$ and finite number of points in time $\{ \tau_1
, \tau_2 , \dots , \tau_{n} \}$ where the state has changed from some $\vert
\alpha_{k-1} \rangle$ to $\vert \alpha_{k} \rangle$.
Within this description the action of the trajectory
is known exactly, since one may formally think about taking the limit 
$\Delta \tau \to 0$ while calculating $S$ within the standard approach.
When we turn to the problem of generating new
kink-antikink pairs, we formally face an incontestable obstacle: the
probability to accept the trajectory with one extra kink-antikink pair is
formally zero, $W_{new}/W_{old}\sim (t\Delta \tau )^2\to 0$, so that the
algorithm does not allow to change the number of kinks at the level of
comparing two particular trajectories. This obstacle, however, can be
circumvented, if one knows an exact {\it integrated} statistics of
trajectories with a given number of kinks. One may decide then about the
number of kinks on the trajectory by relating these integrated
probabilities. In what follows we
demonstrate explicitly how to find this integrated statistics for arbitrary
discrete system in any time window of width $\tau _o$ by relating it to the
statistics of the biased two-level system. 

{\bf 2.} Let $H_0$ and $V$ be the diagonal and non-diagonal parts of the
Hamiltonian, $H$, in a chosen representation corresponding to the full set, 
$\{ \alpha \}$, of eigenstates of $H_0$, and 
$H_0 \mid \alpha \, \rangle = E_{\alpha} \mid \alpha \, \rangle $.
Then statistical operator can be ordinarily related to the Matsubara
evolution operator, $\sigma$, in the interaction representation, i.e., we
write $e^{- \beta H} = e^{- \beta H_0} \sigma $ with 
\begin{equation}
\sigma  =  1 - \int_0^{\beta} d \tau \; V(\tau) + \ldots  
 +  (-1)^m \int_0^{\beta} d \tau_m  \cdots
\int_0^{\tau_2} d \tau_1 \: V(\tau_m) \cdots V(\tau_1) +
\ldots \; ,  \label{rel1}
\end{equation}
where $V(\tau) = e^{ \tau H_0} V e^{- \tau H_0}$. Without loss of generality
and in accordance with typical forms of Hamiltonians of interest, $V$ can be
written as a sum of elementary terms, $Q_s$, whose action on any function
from the set $\{ \alpha \}$ results in another function from this set: 
\begin{equation}
V = \sum_s Q_s \; ;\;\;\;\;
Q_s \mid \alpha \, \rangle = - q_{\gamma \alpha}(s) \mid \gamma \, \rangle
\;\;\;\;\;\;(\gamma = \gamma (s,\alpha )) \;.  \label{Q}
\end{equation}
Since $V$ is Hermitian, for any $s$ in the sum (\ref{Q}) there exists 
$s^{\prime}$ such that $Q_{s^{\prime}} = Q^{\dag}_s$. Then we may rewrite
Eq.\ (\ref{rel1}) in components (below $E_{\alpha \gamma} \equiv E_{\alpha}
- E_{\gamma}$): 
\begin{eqnarray}
\sigma_{\alpha \gamma} & = & \delta_{\alpha \gamma} + \sum_s \int_0^{\beta}
d \tau \: q_{\alpha \gamma}(s) e^{\tau E_{\alpha \gamma}} + \ldots  \nonumber
\\
& + & \sum_{s_1, \ldots , s_m} \int_0^{\beta} d \tau_m \cdots
\int_0^{\tau_2} d \tau_1 \: q_{\alpha \nu}(s_m) e^{\tau_m E_{\alpha \nu}}
\cdots q_{\lambda \gamma}(s_1) e^{\tau_1 E_{\lambda \gamma}} + \ldots \; .
\label{rel2}
\end{eqnarray}
Note that there is no additional summation over the indexes of the
intermediate complete sets (labeled by Greek letters) since these are
defined in a unique way by configurations of $(s_1, s_2,\ldots , s_m)$.

We confine ourselves to the case of finite-range interaction, which is
defined by the requirement that for each term $s_1$ of elementary operators 
$\{ Q_s \}$ there exists only a finite number of terms $s_2$ for which the
condition 
\begin{equation}
\left[ \, Q_{s_1}(\tau_1), \, Q_{s_2}(\tau_2) \, \right] = 0  \label{comm}
\end{equation}
is not met. In the case of finite-range interaction the structure of the
series (\ref{rel2}) is drastically simplified, the simplification being of
crucial importance for practical realization of our algorithm. 
From (\ref{comm}) it follows that up to irrelevant change in indexing energies
and matrix elements, one may pay no attention to the chronological order of 
$Q_{s_1}(\tau_1 )$ and $Q_{s_2} (\tau_2 )$ in the evolution operator. This
suggests representing  of a general
term of the series (\ref{rel2}) in the following form. First, we introduce
the notion of a ``kink of kind $s$", which is characterized by a time 
$\tau$, a matrix element $q_{\alpha \gamma}(s)$, 
and a diagonal-energy difference $E_{\alpha \gamma}$. 
The former two we will refer to as parameters of the
kink. It is essential that (i) to obtain parameters of a kink one need not
know explicitly the whole state $\mid \alpha \, \rangle$, or $\mid \gamma \,
\rangle$ - local information is enough; (ii) to set a particular structure
of a term in Eq.\ (\ref{rel2}), including the chronological order of all
noncommuting operators, it is enough to specify for each kink its associated
neighbors, i.e. nearest in time noncommuting kinks.

Now we are ready to introduce a stochastic process directly evaluating Eq.\ 
(\ref{rel2}). For simplicity, we assume that all $q_{\alpha \beta}(s)$ are
positive real numbers. [In many particular alternative cases a
straightforward generalization is possible, but usually at the expense of
convergence.] Summations and integrations in Eq.\ (\ref{rel2}) then can be
regarded, up to a normalizing factor, as an averaging over a statistics of
different configurations of kinks, each configuration being defined by a
certain number of kinks of certain kinds, their associations and particular
positions in imaginary time. Our aim is to organize a stochastic process of
examining this statistics by generating different kink configurations in
accordance with their weights. The process will consist of a number of
independent sub-processes, responsible for modifications of particular
types. The simplest sub-process is the random shift of the position of a
kink in imaginary time without interchange with associated neighbors. The
weight function for this sub-process is just $\exp (\tau E_{\alpha \gamma})$,
 $\tau$ is the position of the kink, $E_{\alpha \gamma}$ is the
corresponding energy difference. It is clear that this process is far from
being complete since it does not change neither the number of kinks nor
their associations.

The sub-process which is capable of doing so, and which will play the key
part in the whole procedure, is the process of creation/annihilation of a
kink - anti-kink pair. 
First we define the notion of the appropriate window (with respect to
kink - anti-kink pairs of the kind $s$ - $s^{\prime}$), as the interval of
imaginary time of some chosen length $\tau_0$ where (i) there is either one
and only one kink - anti-kink pair of the kind $s$-$s^{\prime}$ (case $A$),
or no such pairs at all (case $B$), and (ii) there are no kinks associated
with any kink of the pair. Now we notice that the total contribution of
configurations of the class $A$ (integrated over the positions of the pair
inside the window and integrated and summed over all possible configurations
of the rest of the kink pattern) differs from the total contribution of the
configurations of the class $B$ by the factor 
\begin{equation}
I_{\alpha \gamma }^{(2)} (s) = \int_{\tau_a}^{\tau_a + \tau_0} d \tau_2
\int_{\tau_a}^{\tau_2} d \tau_1 \; q_{\alpha \gamma}(s^{\prime}) e^{\tau_2
E_{\alpha \gamma}} q_{\gamma \alpha}(s) e^{\tau_1 E_{\gamma \alpha}} \; ,
\label{I1}
\end{equation}
where $\tau_a$ - is a randomly chosen left boundary of the time window.
Elementary integration yields 
\begin{equation}
I_{\alpha \gamma }^{(2)}(s) = \frac{\mid q_{\gamma \alpha}(s) \mid^2}{
E_{\alpha \gamma}^2} \left[ e^{\tau_0 E_{\alpha \gamma}} - \tau_0 E_{\alpha
\gamma} -1 \right] \; .  \label{I2}
\end{equation}
This allows us to weigh these two classes. Having found an appropriate
window we renew its contents by the following rule. First, we delete a pair 
$s$ - $s^{\prime}$, if there is one. Then with the probability $w_A$ we
create another kink - anti-kink pair $s$ - $s^{\prime}$, and with the
probability $w_B = 1-w_A$ leave it as it is. 
\begin{equation}
w_A/w_B = I_{\alpha \gamma}^{(2)}(s) \; .  \label{AB}
\end{equation}
In the former case we also randomly choose the positions of the kink and
anti-kink, in accordance with the distribution following from their
exponential factors $e^{(\tau_2-\tau_1)E_{\alpha \gamma}}$.

It is worthnoting that the kink - anti-kink procedure actually relies on an
important general fact, easily seen from Eq.\ (\ref{rel2}), that the local
statistics of kink - anti-kink chains, corresponding to virtual transitions
between two states, $\mid \alpha \,\rangle $ and $\mid \gamma \,\rangle $,
is exactly the same as for the biased two-level system with the bias energy
equal to $E_{\alpha \gamma }$ and tunneling amplitude equal to $q_{\alpha
\gamma }(s)$. The unnormalized probabilities of having exactly $n$
kink-anti-kink pairs in the appropriate window are given then by 
\begin{equation}
I_{\alpha \gamma }^{(2n)}(s)={\frac{\left( \mid q_{\gamma \alpha }(s)\mid
^2\tau _0^2\right) ^n}{n!}}\:{\frac{\sqrt{\pi }}2}\:e^{-x}\: {\frac{
I_{n-1/2}(x)-I_{n+1/2}(x)}{(2x)^{n-1/2}}}\;,  \label{In}
\end{equation}
where $x=\tau _0E_{\alpha \gamma }/2$, and $I_\nu (x)$ is the modified
Bessel function. Therefore, in principle a more general weighting procedure
is possible, involving transformations of the kink - anti-kink chains of
arbitrarily large length, which first deletes the whole chain of the kink -
anti-kink pairs $s$ - $s^{\prime }$ from the selected appropriate window and
then creates another chain according to Eq.(\ref{In}). This, however, can
hardly have practical importance because of small probability to find an
appropriate window of large length.

Generally speaking, the reason why the kink - anti-kink process in certain
cases turns out to be incomplete is that there may exist more than one way
to get from a state $\mid \alpha \, \rangle$ to a state $\mid \gamma \,
\rangle$. Consider, for example, a typical situation corresponding to the
following two kink chains 
\begin{equation}
q_{\gamma \lambda}(s_2) \, e^{\tau_2 E_{\gamma \lambda}} \; q_{\lambda
\alpha}(s_1) \, e^{\tau_1 E_{\lambda \alpha}} \; \; \; \; 
\mbox{(case $C$)}
\; ,  \label{C}
\end{equation}
\begin{equation}
q_{\gamma \nu}(s_4) \, e^{\tau_4 E_{\gamma \nu}} \; q_{\nu \alpha}(s_3) \,
e^{\tau_3 E_{\nu \alpha}} \; \; \; \; \mbox{(case $D$)}\; .  \label{D}
\end{equation}
To obtain relative weights of classes $C$ and $D$ one could proceed in
direct analogy with the kink - anti-kink processes, comparing the integral
contributions. It is more reasonable, however, to take advantage of the fact
that the number of kinks is the same in both cases. This allows us to take
the differential limit. Namely, we compare only configurations when $\tau_1$
and $\tau_3$ ($\tau_2$ and $\tau_4$) vary in an infinitesimal interval $d\tau
$ around $\tau_1 = \tau_3 = \tau_a$ ($\tau_2 = \tau_4 = \tau_b$). As a
result, we come to a Metropolis-type discrete sub-process of replacing the
pair $(s_1,s_2)$ by the pair $(s_3,s_4)$ and vice versa with probabilities 
$w_D$ and $w_C$ (respectively), satisfying 
\begin{equation}
\frac{w_D}{w_C} \, = \, \frac {q_{\gamma \nu}(s_4) q_{\nu \alpha}(s_3)}
{q_{\gamma \lambda}(s_2) q_{\lambda \alpha}(s_1)} \, \exp \left[ \, (\tau_b
- \tau_a)E_{\lambda \nu} \, \right] \; .  \label{DC}
\end{equation}

All the above-considered sub-processes deal with the ``bulk" of the kink
configuration. Finally, we introduce an ``edge" sub-process, which, on one
hand, changes one of the indexes of $\sigma_{\alpha \gamma}$ (for
definiteness, $\gamma$ into $\lambda$), and, on another hand, changes the
number of kinks by one. The appropriate window for the kink $s$ is now
defined as an interval $[0, \tau_0]$, where there is only one kink 
(case $E$) or zero (case $F$) kinks of the kind $s$, 
and no kinks associated with $s$
. In close analogy with the kink - anti-kink sub-process, in the new
configuration with the probability $w_E$ there will be a kink, and with the
probability $w_F = 1-w_E$ it will be absent 
\begin{equation}
\frac{w_E}{w_F} \, = \, q_{\gamma \lambda}(s) \int_0^{\tau_0} d\tau \,
e^{\tau E_{\gamma \lambda}} \; .  \label{EF}
\end{equation}
The position of the kink in the former case is chosen in accordance with the
factor $e^{\tau E_{\gamma \lambda}}$.

To illustrate the above-described approach, let us formulate it for bosonic
Hubbard models [and closely related to them (by Holstein-Primakoff
transformation) spin lattices]. In the site representation, the diagonal
part of the Hamiltonian, $H_0$, involves interparticle interactions and an
external potential, if any. The nondiagonal part, $V$, corresponds to the
intersite hopping, and terms $Q_s$ describe hopping from one particular site
to another. The most transparent case is a hard-core one- dimensional system,
especially if the quantities of interest are just potential and kinetic
energies. As is easily seen, in this case the kink - anti-kink sub-process
is enough to reproduce the statistics. To enhance the convergence, however,
it is useful to utilize also the sub-process of the random shift of the kink
position and the discrete sub-process of the kink - anti-kink pair reversal,
i.e., Eq.\ (\ref{DC}) with $s_2=s_1^{\prime}$, $s_4=s_1$, $s_3=s_1^{\prime}$. 
In $d \geq 2$ the discrete two-kink sub-process Eq.\ (\ref{DC}) is
essential to reproduce all possible trajectories of a particle. If one is
interested in non-local (both in space and imaginary time) off-diagonal
correlators or, say, the system behavior in a ring geometry with a global
gauge phase, he should take advantage of the edge sub-process 
Eq.\ (\ref{EF}).

It should be noted that the requirement that the interaction be of
finite-range character is not of fundamental nature. In principle our
process is applicable to interactions of arbitrarily long range, and, in
particular, may work in momentum representation - if this is crucial for
this or that reason. The only limitation which arises here is the small size
of the appropriate window $\tau_0$.

We note also, that our approach might prove to be the most effective and
very accurate method for MC simulations of continuous systems if one first
approximate them by lattice models. For one thing, in the case of harmonic
oscillator the corrections for discreteness are known to be extremely
small.

{\bf 3.} All the above mentioned ideas were realized in a code allowing
calculations of the one-particle properties for the Hamiltonian (\ref{4}).
The results were compared with those obtained by the standard Monte Carlo
world-line algorithm, based on finite $\Delta \tau $ \cite{Spasibo}. We
found that, e.g., for the periodic potential $U_i=(-1)^iU$ and discrete
harmonic oscillator $U_i=Ui^2$ the discrete-time code with $\Delta \tau
=0.1/t$ was more then two orders slower in accumulating statistics (or
required $>10^2$ longer computation time in order to achieve the same
accuracy) than ours. This result is hardly surprising, because, as discussed
above, the ``bottleneck'' in the standard procedure is connected with the
unlikely process of the kink-anti-kink pair creation (with probability 
$(t\Delta \tau )^2\sim 10^{-2}$).
No matter how simple are our test models their
description, nevertheless, contains all the essential features of simulation
of much more complicated objects.

We are grateful to V. Kashurnikov for his permanent interest and numerous
discussions. This work was supported by Grant No. INTAS-93-2834-ext [of the
European Community] and partially by the Russian Foundation for Basic
Research [Grant No. 95-02-06191a].

\end{document}